\documentclass[twocolumn,showpacs,superscriptaddress,preprintnumbers,amsmath,amssymb,prc]{revtex4-1}

\usepackage{graphicx}
\usepackage{dcolumn}
\usepackage{bm}
\usepackage{ifpdf}
\usepackage{epstopdf}
\usepackage{slashed}
\usepackage{amsfonts}
\usepackage{mathrsfs}
\usepackage{amssymb}
\usepackage{multirow}
\usepackage{CJK}
\usepackage{color}
\usepackage{subfigure,dcolumn}
\usepackage[T2A,T1]{fontenc}
\usepackage[russian,english]{babel}
\usepackage{amsmath}
\usepackage{listings}
\usepackage{booktabs}
\usepackage{youngtab}
\usepackage{hyperref}

\begin{document}

\title{Bayesian analysis of properties of nuclear matter with the FOPI experimental data}

\author{Guojun Wei}
\affiliation{School of Science, Huzhou University, Huzhou 313000, China}
\affiliation{Institute of Theoretical Physics, Shanxi University, Taiyuan 030006, China}

\author{Manzi Nan}
\affiliation{School of Science, Huzhou University, Huzhou 313000, China}

\affiliation{Institute of Modern Physics, Chinese Academy of Science, Lanzhou 730000, China}

\author{Pengcheng Li}
\affiliation{School of Science, Huzhou University, Huzhou 313000, China}

\author{Yongjia Wang}
\email[Corresponding author, ]{wangyongjia@zjhu.edu.cn}
\affiliation{School of Science, Huzhou University, Huzhou 313000, China}
\affiliation{Guangxi Key Laboratory of Nuclear Physics and Nuclear Technology, Guangxi Normal University, Guilin 541004, China}
\author{Qingfeng Li}
\email[Corresponding author, ]{liqf@zjhu.edu.cn}
\affiliation{School of Science, Huzhou University, Huzhou 313000, China}
%\affiliation{Institute of Modern Physics, Chinese Academy of Science, Lanzhou 730000, China}
\author{ Gaochan Yong}
\affiliation{Institute of Modern Physics, Chinese Academy of Science, Lanzhou 730000, China}
\author{Fuhu Liu}
%\email{fuhuliu@sxu.edu.cn}
\affiliation{Institute of Theoretical Physics, Shanxi University, Taiyuan 030006, China}
\date{\today}

\begin{abstract}
Based on the ultra-relativistic quantum molecular dynamics (UrQMD) transport model, combined with experimental data of directed flow, elliptic flow, and nuclear stopping power measured by FOPI in $\rm ^{197}Au+^{197}Au$ collisions at beam energies ($E_{lab}$) of 0.25 and 0.4 GeV/nucleon, the incompressibility of the nuclear equation of state $K_0$, the nucleon effective mass $m^*$, and the in-medium correction factor ($F$, with respect to free-space values) on the nucleon-nucleon elastic cross sections are studied by Bayesian analysis. It is found that both $m^*$ and $F$ can be tightly constrained with the uncertainty $\le$ 15\%, however, $K_0$ cannot be constrained tightly. We deduce $m^*/m_0 = 0.78^{+0.09}_{-0.10}$ and $F = 0.75^{+0.08}_{-0.07}$ with experimental data at $E_{lab}$ = 0.25 GeV/nucleon, and the obtained values increased to $m^*/m_0 = 0.88^{+0.03}_{-0.03}$ and $F = 0.88^{+0.06}_{-0.07}$ at $E_{lab}$ = 0.4 GeV/nucleon. The obtained results are further verified with rapidity-dependent flow data.

\end{abstract}

\maketitle

%%%%%%%%%%%%%%%%%%%%%%%%%%%%%%%%%%%%%%%%%%%%%%%%%%%%%%%%%%%%%%%%%%%%%%%%%%%%%%%%%%%%%%%%%%%%%
\section{Introduction}\label{sec:1}
The nuclear equation of state (EoS) is a fundamental relationship that describes how nuclear binding energy depends on variables such as temperature, density and isospin asymmetry. The information on the nuclear EOS is crucial for advancing our knowledge of the nuclear force and the boundaries of the chart of nuclei, exploring the structure of dense stars and exotic nuclei, and validating existing nuclear theoretical models \cite{Drischler:2021kxf,xu2019transport,li2008recent,Sorensen:2023zkk,Huth:2021bsp,SpiRIT:2023htl,Tsang:2023vhh,doi:10.1126/science.1078070}.

The nuclear EoS can be deduced by comparing measurements in experiments with simulations of theoretical models. Over the past decades, significant progress has been made in constraining the nuclear EoS at densities below and near the saturation density. However, the whole picture of the nuclear EoS within a broad range of density is still missing, sometimes constraints
obtained from different observables or from different theoretical models lead to different conclusions \cite{TMEP:2022xjg}. For example, the incompressibility of isospin-symmetric nuclear matter $K_0$, which is one of the key parameters to describe the EoS, was extracted to be approximately 220 MeV from isoscalar giant monopole resonances data in
comparison with theoretical calculations \cite{garg2018compression,xu2021bayesian,li2023toward}. To best reproduce the production rate of $K^+$ mesons in heavy-ion collisions (HICs) around 1 GeV/nucleon incident energy, a value of around 200 MeV was deduced by simulations of transport models \cite{fuchs2001probing,hartnack2006hadronic,sturm2001evidence,Fuchs:2000kp}. Using FOPI data on the rapidity-dependent elliptic flow of protons and deuterons in $\rm ^{197}Au+^{197}Au$ collisions between 0.4 and 1.5 GeV/nucleon, the incompressibility $K_0$ was determined to be 190$\pm$30 MeV in the analysis with the isospin quantum molecular dynamics (IQMD) model \cite{LeFevre:2015paj}. While $K_0$ = 220$\pm$40 MeV was extracted with the same FOPI data by using the ultra-relativistic quantum molecular dynamics (UrQMD) model \cite{Wang:2018hsw}. When going to higher densities, the situation becomes more complicated, even though observables in HICs at higher (e.g., RHIC-BES) energies and astrophysical observations (e.g., the properties of neutron stars and the gravitational-wave) can be used to infer the EoS \cite{Oertel:2016bki,Pang:2022rzc,Koehn:2024set,Zhang:2024npg,
Chatziioannou:2024tjq,Semposki:2024vnp,Sammarruca:2024rvc}. 
There are a few possible reasons for the observed deviations among
different studies: (I) different assumptions and coding techniques are used in different theoretical models \cite{SpiRIT:2020sfn,TMEP:2016tup,TMEP:2017mex,TMEP:2019yci,TMEP:2021ljz,TMEP:2022xjg}; (II) different observables may reflect the properties
of nuclear EOS at different density regions \cite{Gao:2022shr,Wang:2024ktk,liu2023impacts,Tsang:2023vhh}; (III) different techniques and methods for data processing are used in obtaining observables in different experiment \cite{Andronic:2006ra,HADES:2020ver,Mamamev:2024ynl,Sharma:2023tvv}.

Besides, the nucleon effective mass ($m^*$) and the in-medium nucleon-nucleon elastic cross section (NNECS) together with $K_0$ are important ingredients of transport model to simulate HICs at a few hundred MeV per nucleon \cite{xu2019transport,ono2019dynamics,TMEP:2022xjg,bleicher2022modelling}. $m^*$ characterizes the momentum-dependence of the nuclear potential, and the in-medium NNECS arises due to the dressing, which refers to the
modification of a particle line with loop diagrams. A modification
factor $F=\sigma_{NN}^{medium}/\sigma_{NN}^{{free}}$, which is defined as the ratio of the cross section in the nuclear medium $\sigma_{NN}^{medium}$ to that in the free space $\sigma_{NN}^{free}$, is used to characterize how strong this modification is. Usually, $F$ is regarded as density, asymmetry and momentum-dependent. Here, as a simple approximation it is assumed as an effective constant, valid in the range of densities and momenta attained at the energies considered in the present work. As will be discussed later, this approximation is probably not sufficient. At present, the constraints on $m^*$ and $F$ are still debated \cite{li2008recent,li2018nucleon,Han:2022quc,Cui:2018dex,SpiRIT:2023htl,song2023medium,li2023bayesian,li2025evolutions}. 

Bayesian analysis is a statistical method that can be used to extract information from the comparison between experimental data and theoretical calculations. It provides a consistent method for the extraction of
parameters from multiple experimental observables, and yields results with quantified uncertainties \cite{gal2022bayesian}. Bayesian analysis has been widely used in heavy-ion physics, to extract the specific shear and bulk viscosity of quark–gluon plasma from ALICE experimental data, to constrain the density-dependent nuclear symmetry energy and the
density dependence of the EoS \cite{li2023bayesian,SpiRIT:2023htl,li2025evolutions,Tsang:2023vhh,PhysRevC.110.064911}. In a recent work, by using experimental data accumulated in HICs at center-of-mass energy ranging from $\sqrt{s_{NN}}$ = 2.24 to 8.86 GeV, the EoS of dense matter in the range 2-4 times the nuclear saturation density ($\rho_0$) was studied with Bayesian analysis \cite{kuttan2022qcd}. Indeed, Bayesian analysis of properties of nuclear matter around $\rho_0$ with 
experimental data in HICs at lower energies is also crucial in determining the EoS of dense matter within a broad range of densities, and in testing calculations of theoretical models and results deduced from nuclear structure and neutron star data \cite{Huth:2021bsp,Bernhard:2018hnz}. In the present work, $K_0$, $m^*$, and $F$ are studied by Bayesian analysis using FOPI experimental data, and the results are compared with those from Refs. \cite{PhysRevC.110.064911,li2025evolutions,Wang:2018hsw,wang2024bayesian}

%The aim of the present work is to constrain the nuclear incompressibility $K_0$, the nucleon effective mass $m^*$, and the in-medium correction factor $F$ on the nucleon-nucleon elastic cross section using Bayesian analysis of FOPI experimental data

The paper is organized as follows. We describe the experimental observables and transport model simulations in Sec. \ref{sec:2}. We constrain $K_0$, $m^*$ and $F$ by Bayesian analysis and validate the constrained results
with rapidity-dependent flow in Sec. \ref{sec:3}. Finally, Sec. \ref{sec:4} contains a summary and an outlook.

\section{Experimental Observables and transport model simulations}\label{sec:2}

%%%%%%%%%%%%%%%%%%%%%%%%%%%%%%%%%%%%%%%%%%%%%%%%%%%%%%%%%%%%%%%%%%%%%%%%%%%%%%%%%%%%%%
%\subsection{observable selection}
The collective flow describes the collective motion of particles produced in HICs and is of great significance to the study of many topics in HICs across a wide range of energy \cite{reisdorf1997collective,herrmann1999collective,doi:10.1126/science.1078070,heinz2013collective,Lan2022,elfner2023exploration,Wang:2020dru}. The directed flow $v_1$ and the elliptic flow $v_2$ are two components widely studied. They can be obtained by Fourier expansion of the azimuthal distribution of the particles:

\begin{equation}
v_{1} \equiv\langle\cos (\phi)\rangle=\left\langle\frac{p_{x}}{p_{t}}\right\rangle,
\end{equation}

\begin{equation}
v_{2} \equiv\langle\cos (2 \phi)\rangle=\left\langle\frac{p_{x}^2-p_{y}^2}{p_{t}^2}\right\rangle.
\end{equation}
Here $p_x$ and $p_y$ are two components of the transverse momentum $p_t=\sqrt{p_{x}^{2}+p_{y}^{2} } $. Angle brackets represent the average of all the selected particles from all events. The directed flow $v_1$ reflects the motion of particles in
the reaction plane (defined by the impact parameter $b$ in
the $x$-axis and the beam direction in the $z$-axis), while the elliptic
flow $v_2$ characterizes the motion in the out-of-plane direction. Both $v_1$ and $v_2$ have complex dependence on longitudinal and transverse 
momentum. They are functions of the transverse momentum $p_t$ and rapidity $y_z$ for a certain species of particles produced from HICs with fixed colliding system, beam energy, and impact parameter. Usually, for studying HICs at intermediate energies, the scaled units $y_0$ and $u_{t0}$ are used instead of $y_z$ and $p_t$, a more detailed definition was given in the experimental report \cite{FOPI:2011aa}. Nuclear stopping power describes the efficiency of converting the beam energy in the longitudinal (beam) direction into the transverse direction. A few different observables have been used to measure the nuclear stopping power in HICs, in the present work, the quantity $vartl$ is used. It is defined as the ratio of variances of the transverse to longitudinal rapidity distribution, which reads as:

%In addition to the collective flow, nuclear stopping power is another frequently used observable in HICs. $Vartl$ is a frequently used observable representing nuclear stopping power, defined as the ratio of the variance of the particle velocity along the transverse distribution to the variance of the longitudinal velocity distribution:

%\begin{equation}
%vartl=\frac{\left\langle y_{x}^{2}\right\rangle}{\left\langle y_{z}^{2}\right\rangle}=\frac{\sum N_{x} y_{x}^{2}}{\sum N_{x}} \large/ \frac{\sum N_{z} y_{z}^{2}}{\sum N_{z}}.
%\end{equation}

\begin{equation}
\mathrm{var}tl = \frac{\langle y_x^2 \rangle}{\langle y_z^2 \rangle} = \frac{\sum_{i = 1}^{n} N_{x,i} y_{x,i}^2 / \sum_{i = 1}^{n} N_{x,i}}{\sum_{i = 1}^{n} N_{z,i} y_{z,i}^2 / \sum_{i = 1}^{n} N_{z,i}}
\end{equation}

The $\langle y_x^2 \rangle$ and $\langle y_z^2 \rangle$ are the variances of the rapidity distribution of the particle in the $x$ direction and $z$ direction, respectively; the $N_x$ and $N_z$ denote the numbers of particles in each $y_x$ and $y_z$ bins, and $n$ denote the number of bins \cite{PhysRevLett.92.232301,lehaut2010study,Lopez:2014dga}.

%\subsection{Experimental setup}
The experimental data utilized in this study were the observables of free protons produced in $\rm ^{197}Au+^{197}Au$ collisions at beam energies ($E_{lab}$) of 0.25 and 0.4 GeV/nucleon, measured by the FOPI Collaboration. Detail of the used experimental data are listed in Tab. ~\ref{table:1}. The reduced rapidity $y_0$ distributions of $v_1$ and $v_2$  with $y_0$ < |0.5| and transverse 4-velocities $u_{t0}$ cut were fitted using the functions $v_1$ = $v_{11}$ · $y_0$ + $v_{13}$ · $y_0^3$ and $v_2$ = $v_{20}$ + $v_{22}$ · $y_0^2$ + $v_{24}$ · $y_0^4$, respectively. $v_{11}$ is the slope of the directed flow and $v_{20}$ is the elliptic flow at mid-rapidity ($y_0$ = 0). These two quantities have been widely studied in HICs. In addition, as reported in Refs. \cite{LeFevre:2015paj, Wang:2018hsw}, the quantity
$v_{2n}$ defined by $v_{2n}$  = |$v_{20}$| + |$v_{22}$| is quite sensitive to the incompressibility $K_0$ and the in-medium
nucleon–nucleon elastic cross section, thus $v_{2n}$ is used to validate the inferred results from Bayesian analysis.

%The experimental data selected for this work are from the large acceptance apparatus FOPI experiment\cite{reisdorf2012systematics}, as shown in Tab.~\ref{table:1} and Tab.~\ref{table:2}. The reduced rapidity $y_0$ distributions of $v_1$ and $v_2$ are fitted by $v_1$ = $v_{11}$ · $y_0$ + $v_{13}$ · $y_0^3$ and $v_2$ = $v_{20}$ + $v_{22}$ · $y_0^2$ + $v_{24}$ · $y_0^4$\cite{reisdorf1997collective,lehaut2010study,li2022accessing}. 

\begin{table}[!htbp]
\centering
\caption{List of observables used in the present work. Data are taken from Refs. \cite{PhysRevLett.92.232301,FOPI:2010xrt,FOPI:2011aa}. The reduced impact parameter $b_0$ is defined as
$b_0=b/b_{max}$ with $b_{max} = 1.15 (A_{P}^{1/3} + A_{T}^{1/3})$~fm. }  % 添加标题
\label{tab:tableTab} % 设置标签
\begin{tabular}{ccccc} 
\toprule 
$E_{lab}$ & Observable & impact parameter & $u_{t0}$ cut & value  \\
\midrule 
0.25 & $v_{11}$ & $b_0$ < 0.25 & $u_{t0}$ > 0.8 & 0.23$\pm$0.01  \\
 &  & 0.25 < $b_0$ < 0.45 & $u_{t0}$ > 0.8 & 0.37$\pm$0.01  \\
 \midrule 
0.25 & $-v_{20}$ & $b_0$ < 0.25 & $u_{t0}$ > 0.8 & 0.026$\pm$0.001  \\
 &  & 0.25 < $b_0$ < 0.45 & $u_{t0}$ > 0.8 & 0.046$\pm$0.005  \\
 &  & 0.45 < $b_0$ < 0.55 & $u_{t0}$ > 0.8 & 0.064$\pm$0.002  \\
 \midrule 
0.25 & $vartl$ & $b_0$ < 0.15 & --- & 0.891$\pm$0.041    \\

\midrule 
0.4 & $v_{11}$ & $b_0$ < 0.25 & $u_{t0}$ > 0.8 & 0.31$\pm$0.01  \\
 &  & 0.25 < $b_0$ < 0.45 & $u_{t0}$ > 0.8 & 0.45$\pm$0.02  \\
 &  & $b_0$ < 0.15 & $u_{t0}$ > 0.4 & 0.23$\pm$0.01  \\
 &  & $b_0$ < 0.25 & $u_{t0}$ > 0.4 & 0.28$\pm$0.01  \\
 &  & 0.25 < $b_0$ < 0.45 & $u_{t0}$ > 0.4 & 0.39$\pm$0.02  \\
 &  & 0.45 < $b_0$ < 0.55 & $u_{t0}$ > 0.4 & 0.43$\pm$0.01  \\
\midrule
0.4 & $-v_{20}$ & $b_0$ < 0.25 & $u_{t0}$ > 0.8 & 0.041$\pm$0.002  \\
 &  & 0.25 < $b_0$ < 0.45 & $u_{t0}$ > 0.8 & 0.075$\pm$0.005  \\
 &  & 0.45 < $b_0$ < 0.55 & $u_{t0}$ > 0.8 & 0.111$\pm$0.002  \\
 &  & 0.25 < $b_0$ < 0.45 & $u_{t0}$ > 0.4 & 0.054$\pm$0.003  \\
\midrule
0.4 & $vartl$ & $b_0$ < 0.15 & --- & 0.955$\pm$0.045    \\

\bottomrule 
\end{tabular}
% 标题和标签写在这里也可以
\label{table:1}
\end{table}

In the present work, the version of the UrQMD model modified by the Huzhou group is used. The main modifications are as follows. (I) The initializations of projectile and target nuclei are selected by considering their binding energies. (II) The Skyrme type nucleonic potential energy density is used to determine parameters in the nuclear potential, and the momentum-dependent potential is added. (III) Medium modified elastic nucleon-nucleon cross section (NNCS) is considered instead of the NNCS in the free space. (IV) An isospin-dependent minimum span tree (iso-MST) algorithm is used to construct clusters. Details about the above modifications can be found in Refs. \cite{Li11,wang2014collective,wang2020application,liu2023impacts,Li:2018wpv}. 

\renewcommand{\figurename}{Fig.}
\begin{figure}[b]
    \centering
    \includegraphics[width=0.5\textwidth]{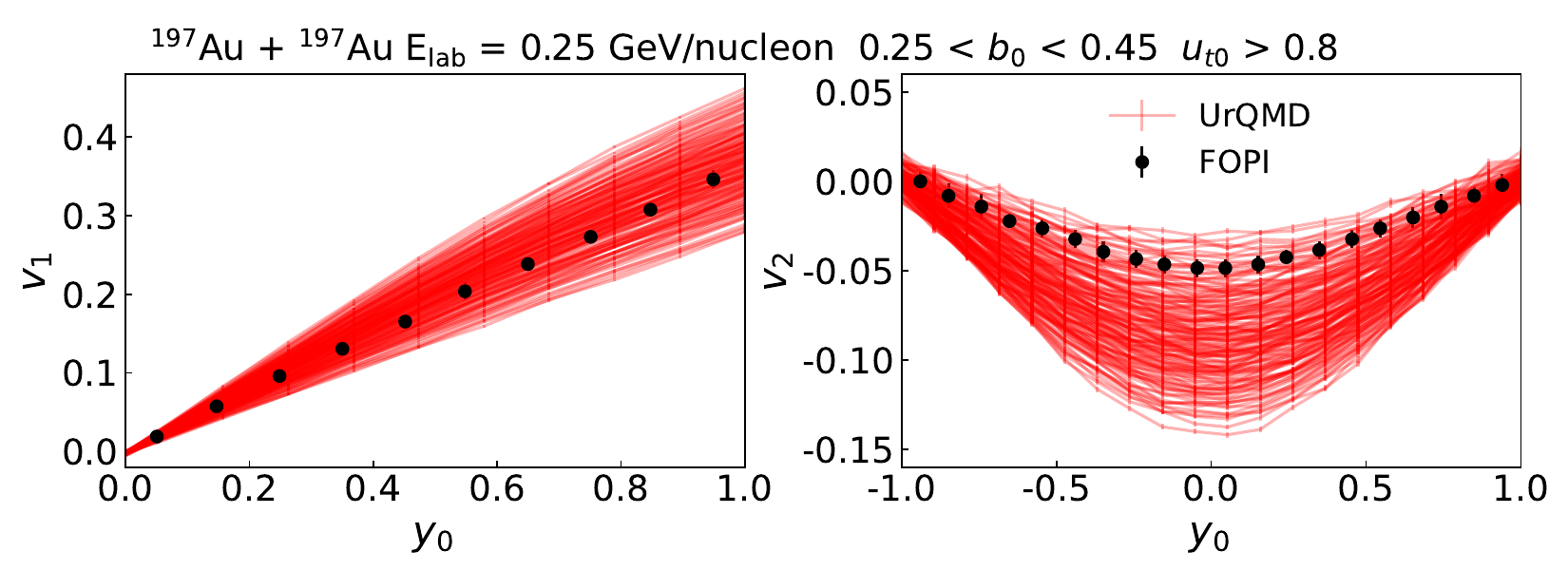}
    \caption{Comparison of UrQMD model data with experimental data. Each red line represents the rapidity distribution of the directed and elliptic flow of free protons from $^{197}$Au+$^{197}$Au collisions at $E_{lab}$ = 0.25 GeV/nucleon, obtained using UrQMD model. The black dots represent the FOPI experimental data taken from Ref. \cite{FOPI:2011aa}.}
    \label{fig:1}
\end{figure}

By varying parameters in the potential term and the in-medium factor $F$, and running the UrQMD model, one can get the simulated results on observables under different $K_0$, $m^*$, and $F$ values. 150 parameter sets of the UrQMD model with $K_0$ = 180, 220, 260, 300, 340, 380 MeV, $m^*/m_0$ = 0.6, 0.7, 0.8, 0.9, 0.95 and $F$ = 0.6, 0.7, 0.8, 0.9, 1.0 are run. 
%Details about how to calculate these parameters are given in the Appendix. 
For each case, 4$\times$10$^5$ events with the reduced impact parameter $b_0$<0.55 are simulated in order to make sure that the statistical errors are negligible. We set intentionally large ranges for each parameter, as shown in Fig. \ref{fig:1}. The experimental data can be fully covered by UrQMD calculations. To do Bayesian analysis, one needs a model emulator to predict the output of the UrQMD model in much less time than an explicit calculation. More detailed information on the use of Bayesian analysis and emulator can be found in the Appendix.

\section{Results and discussions}\label{sec:3}
\begin{figure}[!htbp]
    \centering
    \includegraphics[width=0.5\textwidth]{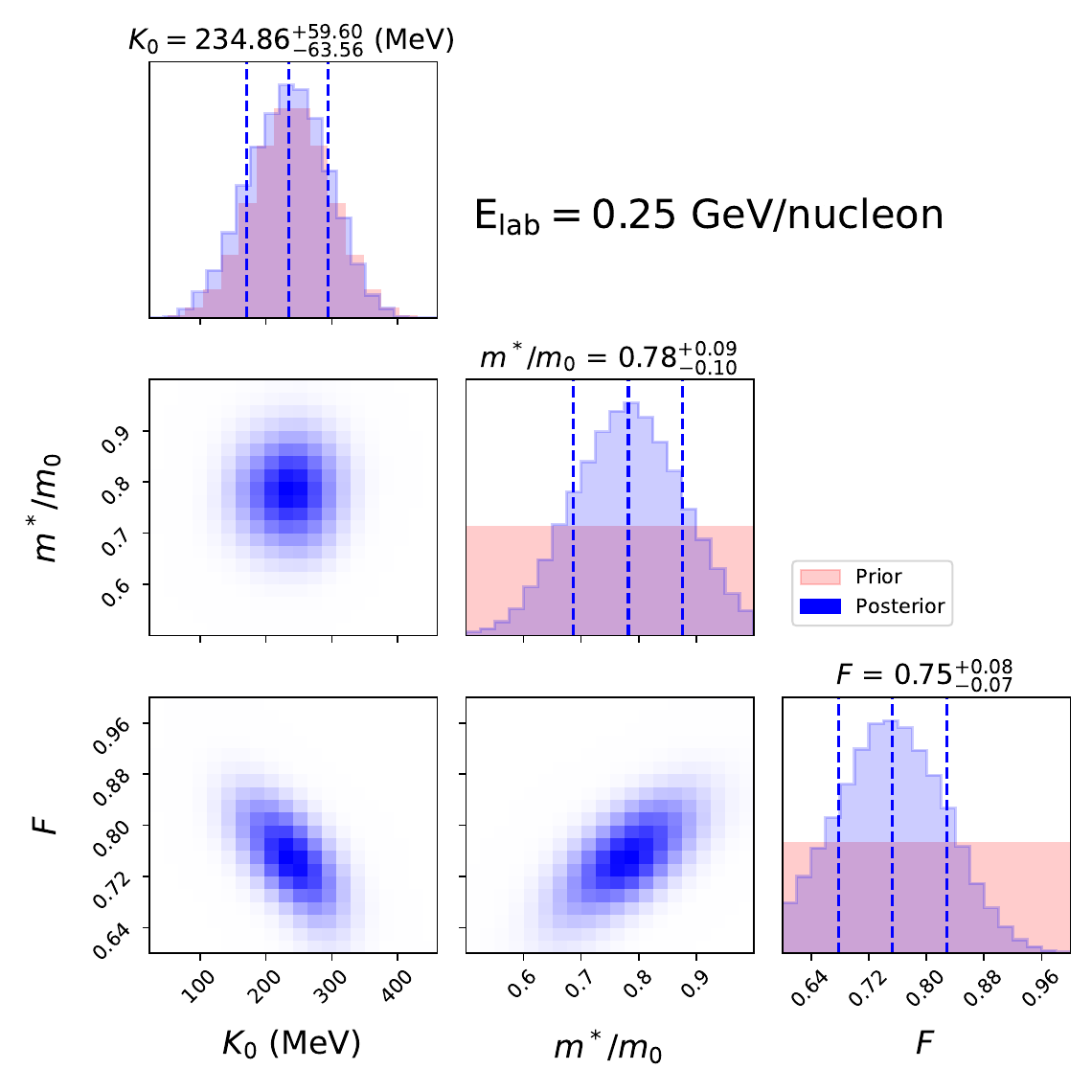}
    \caption{Posterior distributions for the EoS parameters from calibrating at $E_{lab}$ = 0.25 GeV/nucleon. The diagonal has marginal distributions for each parameter, with vertical dashed lines indicating the median and the bounds of the 68\% confidence interval, while the off-diagonal contains joint distributions showing correlations among pairs of parameters.}
    \label{fig:4}
\end{figure}

\begin{figure}[!htbp]
    \centering
    \includegraphics[width=0.5\textwidth]{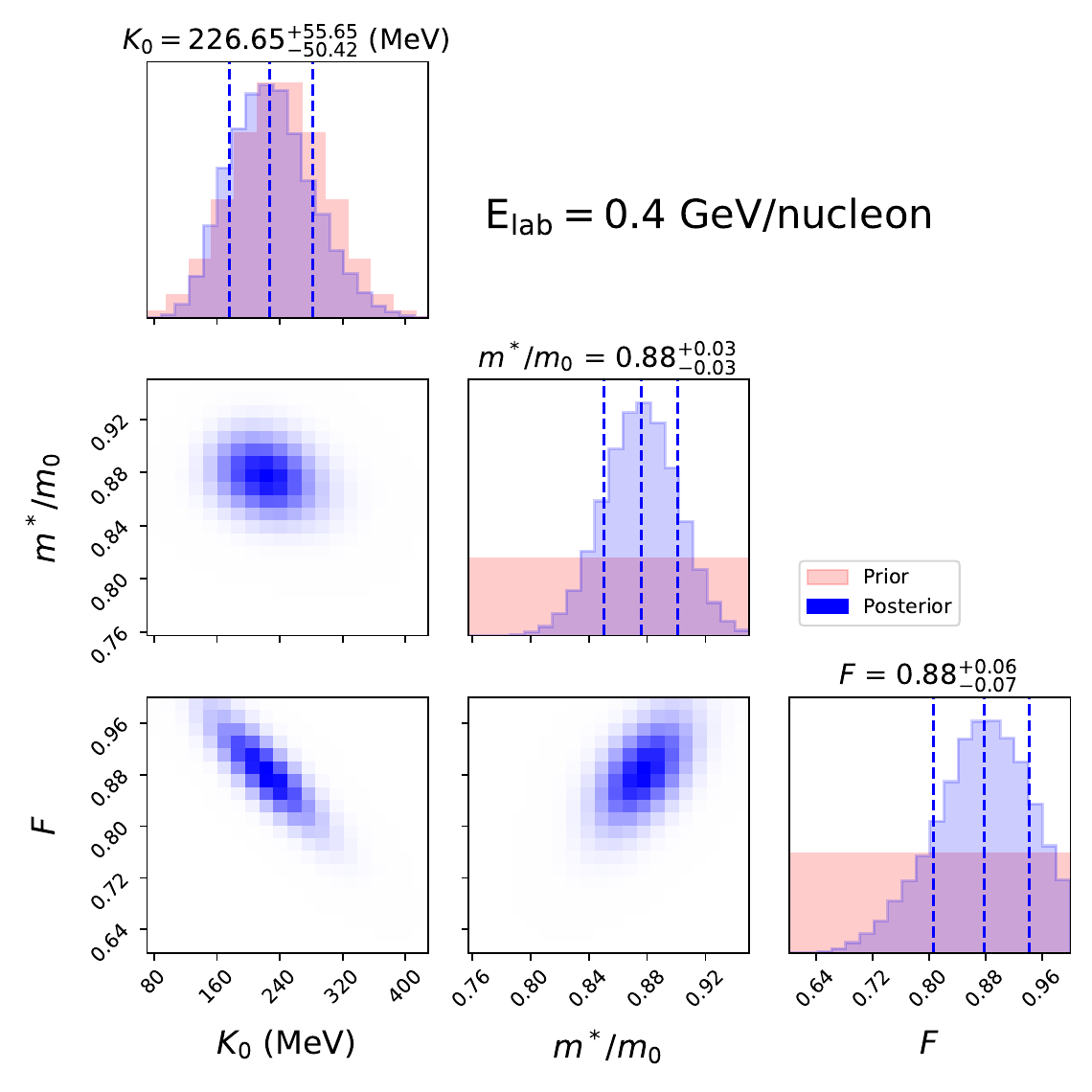}
    \caption{The same as Fig. \ref{fig:4} but for the results constrained with the experimental data at $E_{lab}$ = 0.4 GeV/nucleon.}
    \label{fig:6}
\end{figure}

In this work, the prior distribution of the incompressibility $K_0$ is assumed to be a Gaussian distribution centered at 240 MeV with a width of 60 MeV, and the prior distributions of the nucleon effective mass $m^*$ and the in-medium correction factor $F$ are assumed to be uniform within the model space. Figs. ~\ref{fig:4} and ~\ref{fig:6} present the obtained results with experimental data at $E_{lab}$ = 0.25 and 0.4 GeV/nucleon, respectively. They are the marginal and joint probability distributions for the parameters $K_0$, $m^*$, and $F$. At both beam energies, we see both $m^*$ and $F$ have an approximately narrow normal distribution. This means that both $m^*$ and $F$ are tightly constrained by the FOPI experimental data. At $E_{lab}$ = 0.25 GeV/nucleon, the estimated values for $m^*$/$m_0$ and $F$ are $0.78^{+0.09}_{-0.10}$ and $0.75^{+0.08}_{-0.07}$, respectively, with their 1$\sigma$ intervals similarly represented by the percentiles.
The constrained values are increased to $m^*/m_0 = 0.88^{+0.03}_{-0.03}$ and $F = 0.88^{+0.06}_{-0.07}$ at $E_{lab}$ = 0.4 GeV/nucleon. These central values are consistent with many previous studies, see e.g., Refs. \cite{danielewicz2000determination,Li:2018wpv,li2022accessing,PhysRevC.110.064911,li2025evolutions}. In addition, the increase of $m^*$ and $F$ at higher beam energy are also in line with our previous work \cite{li2022accessing} and recent works done in \cite{song2023medium,li2025evolutions}.
In contrast to $m^*$ and $F$, the posterior distribution of $K_0$ is quite close to its prior distribution. This means that $K_0$ is not tightly constrained in the present work. However, the posterior distribution of $K_0$ has a central value of around 230 MeV at both $E_{lab}$ = 0.25 and 0.4 GeV/nucleon, which tends to give a soft EoS consistent with many previous studies; see, e.g., Refs \cite{hartnack2012strangeness,le2018origin,Du:2023ype,PhysRevC.110.064911}. We have attempted to use a uniform prior for $K_0$ = 180-380 MeV in Bayesian analysis. It is found that both $m^*$ and $F$ can be constrained to Gaussian distributions, and the central values are similar to the results presented in Figs. ~\ref{fig:4} and ~\ref{fig:6}. However, the posterior distribution of $K_0$ is not Gaussian distribution. This is because at the studied beam energies, the influence of $K_0$ on the observables is smaller than that of $m^*$ and $F$. Considering the fact that many current studies tend to give a soft EOS, such as Refs. \cite{LeFevre:2015paj,li2025evolutions,Wang:2018hsw,PhysRevC.110.064911,Kireyeu:2024hjo}, a Gaussian prior for $K_0$ is adopted.

At both beam energies, it can be seen that the correlation between $F$ and $K_0$ is negative. This is due to the fact that both the nuclear stopping power and the strength of the collective flow increase with increasing the stiffness of the nuclear EoS or with increasing $F$, as has been studied in Refs. \cite{Li:2018bus,Li:2018wpv,li2025evolutions}. Accordingly, calculations using either a smaller value of $F$ coupled with a larger value of $K_0$ or the opposite case may be identical. A positive correlation between $F$ and $m^*$ is observed. This is understandable, for example, a smaller $m^*$ which means a stronger momentum-dependent potential will lead to a stronger collective flow \cite{danielewicz2000determination}, while a larger value of $F$ which means a more violent nucleon-nucleon collision may also lead to a stronger collective flow.
It is interesting to note that this positive correlation between $F$ and $m^*$ is in line with the effective mass scaling model for
in-medium NN cross sections \cite{li2008recent,li2018nucleon}. 
%i.e., $F=(\mu_{NN}^{*}/\mu_{NN})^2$ with $\mu_{NN}$ and $\mu_{NN}^{*}$ are, respectively, the free-space and in-medium reduced masses of the colliding nucleon pair. 
Several previous studies have found that some observables calculated with a soft momentum-dependent EoS and with a stiff momentum-independent EoS are degenerate \cite{barker2019shear,isse2005mean,andronic2003directed,Kireyeu:2024hjo}. In this context, a positive correlation between $m^*$ and $K_0$ may be expected. However, the correlations between $m^*$ and $K_0$ shown in Figs. ~\ref{fig:4} and ~\ref{fig:6} are not pronounced. This may be explained as follows. Observables from HICs at different impact parameters are used, and the effect of $m^*$ on observables strongly depends on the impact parameter (usually the effect is more distinct at a larger impact parameter) \cite{danielewicz2000determination}. Thus, the combination of observables from different impact parameters may diminish the correlation between constrained $m^*$ and $K_0$.

\begin{figure*}[!htbp]
    \centering
    \includegraphics[width=\textwidth]{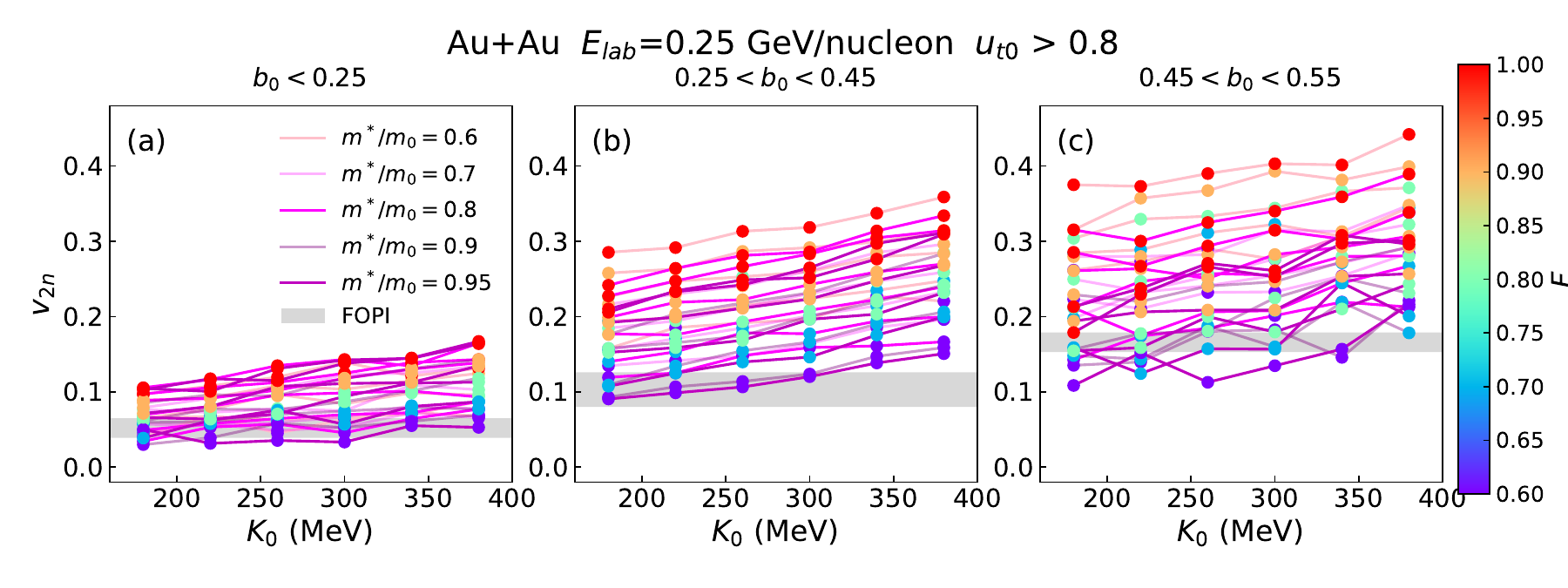}
    \caption{The $v_{2n}$ of free protons produced from $^{197}$Au+$^{197}$Au collisions with different centralities at $E_{\rm lab}$ = 0.25 ~GeV/nucleon are shown as a function for $K_0$, $m^*$ and $F$. In each panel, the shaded bands indicate the FOPI experimental data and full circles denote the UrQMD calculations using different parameter sets. Each dot represents result of UrQMD calculated with a grid of values. The color of dots denote the magnitude of the in-medium correction factor $F$ (see color bar) and different colors of lines denote different $m^*$. The FOPI data are taken from Ref. \cite{FOPI:2011aa}.}
    \label{fig:8}
\end{figure*}

\begin{figure*}[!htbp]
    \centering
    \includegraphics[width=0.7\textwidth]{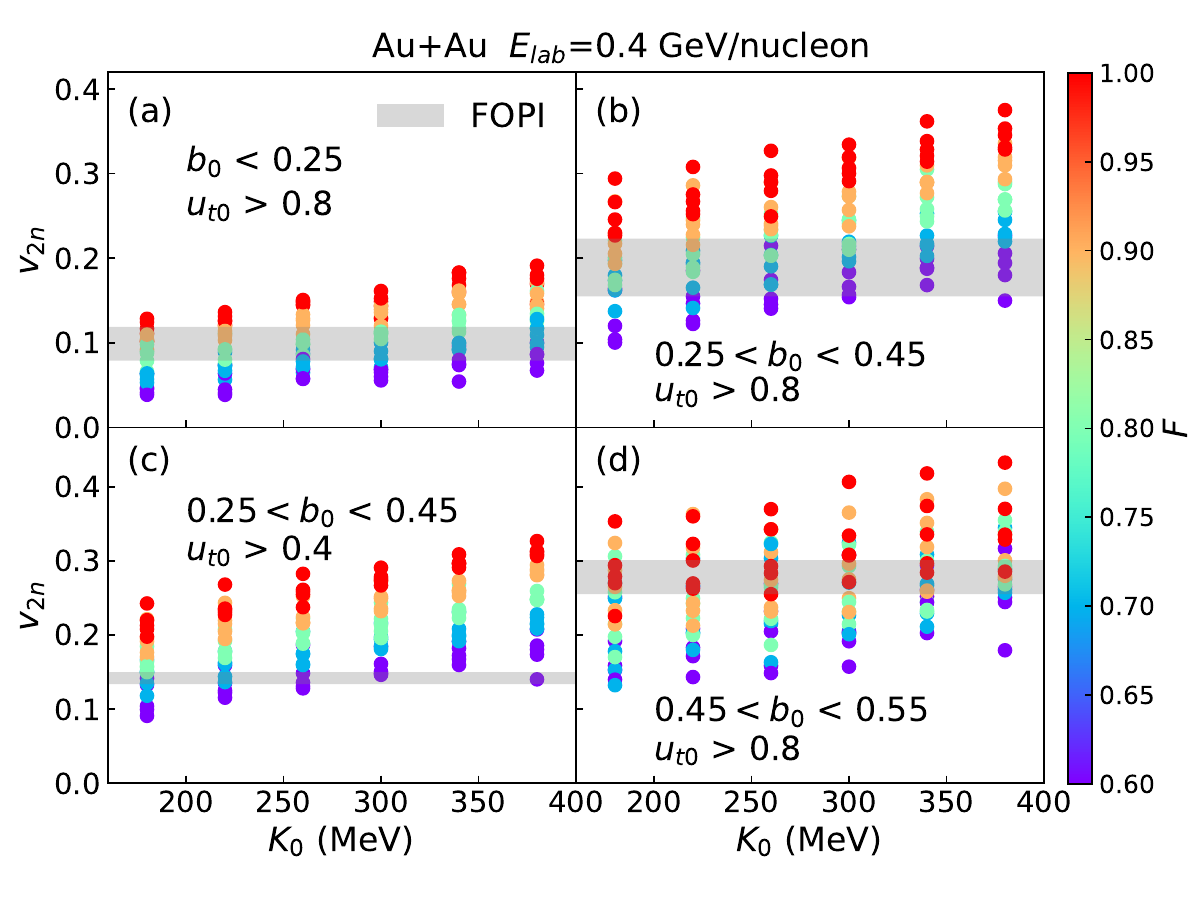}
    \caption{The same as Fig. \ref{fig:8} but for the results at $E_{lab}$ = 0.4 GeV/nucleon. The FOPI data are taken from Ref. \cite{FOPI:2011aa}.}
    \label{fig:9}
\end{figure*}

To further validate the constrained results, the rapidity-dependent elliptic flow parameter $v_{2n}$ that is not used for Bayesian analysis is calculated. The $v_{2n}$ of free protons produced in $^{197}$Au+$^{197}$Au collisions at $E_{\rm lab}$ = 0.25 and 0.4 GeV/nucleon are shown in Figs. \ref{fig:8} and \ref{fig:9}, respectively. First, it can be seen that $v_{2n}$ increases with increasing $K_0$. This manifests $v_{2n}$ is sensitive to $K_0$. However, $v_{2n}$ is also related to the in-medium correction factor $F$ and the nucleon effective mass $m^*$. For central ($b_0$<0.25) and semi-central (0.25<$b_0$<0.45) collisions, the value of $v_{2n}$ is gradually enhanced with increasing $F$, which indicates that the effect of the nucleon effective mass $m^*$ is negligible. However, for peripheral (0.45<$b_0$<0.55) collisions, the value of $v_{2n}$ does not smoothly increase with $F$, which suggests the importance of the nucleon effective mass $m^*$ on $v_{2n}$. Second, in all figures, $v_{2n}$ calculated with simultaneously larger values of $K_0$ and $F$ are far from the FOPI experimental data. Calculations with the central values of the posterior distributions are in line with the experimental data. This means that the posterior distributions extracted from Bayesian analysis are sensible.

\begin{figure*}[!htbp]
   \centering
    \includegraphics[width=0.85\textwidth]{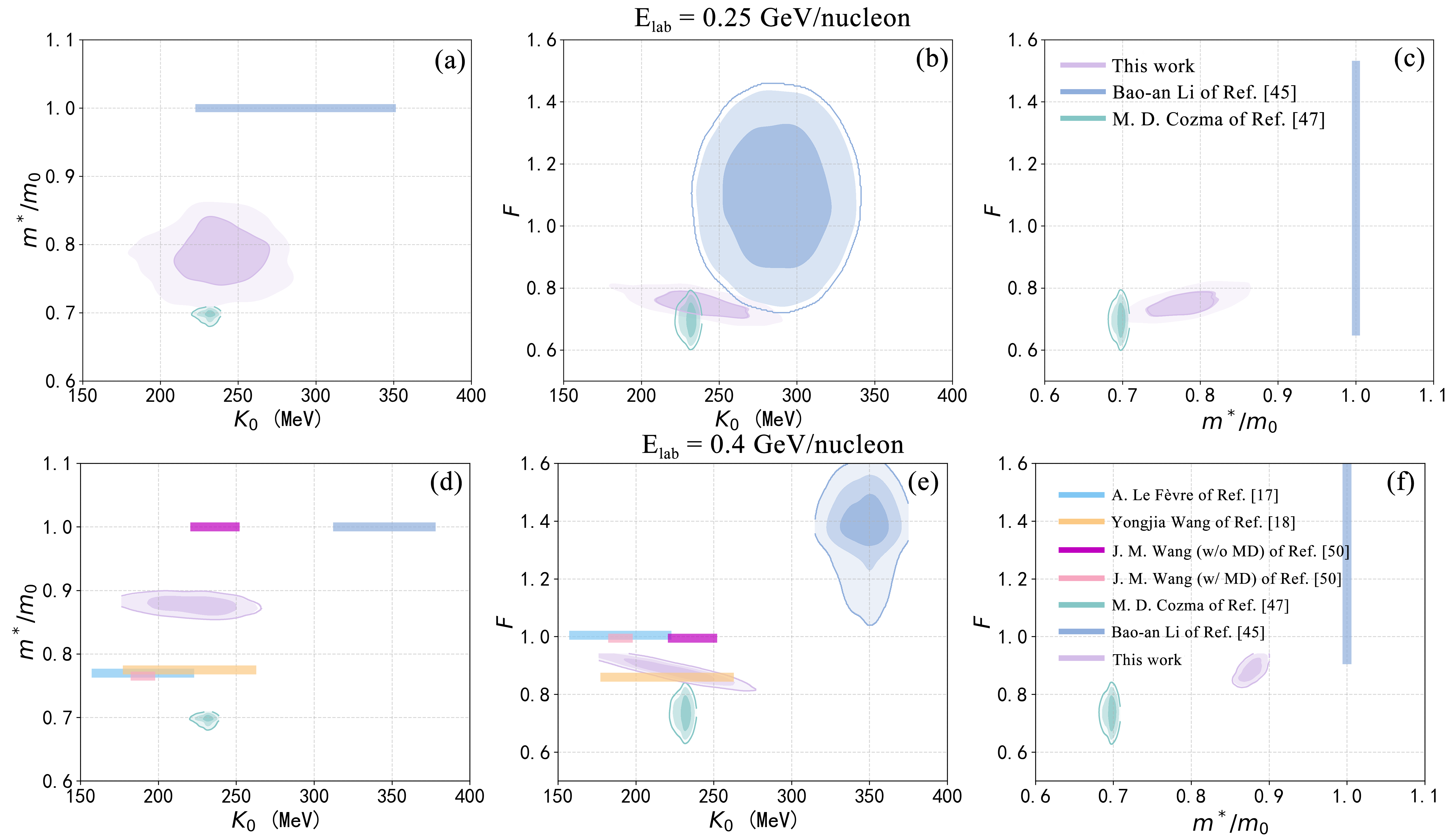}
    \caption{Constraints on the $m^*/m_0$-$K_0$ (a and d), $F$-$K_0$ (b and e), and $F$-$m^*/m_0$ (c and f) correlations obtained from the present work and literature \cite{PhysRevC.110.064911,li2025evolutions,wang2024bayesian}. The filled bands of Bao-An Li and M. D. Cozma are produced through random sampling within the stated uncertainty ranges. In the works of A. Le Fevre $et$ $al.$ and J. M. Wang $et$ $al.$, the free nucleon-nucleon cross section was used, thus $F$ = 1. In the work of Bao-an Li $et$ $al.$, a momentum-independent potential was used, thus $m^*/m_0$ = 1.}
   \label{fig:7}
\end{figure*}
In addition, we collected the results in literature and compared them with those of this work, as shown in Fig. ~\ref{fig:7}. In the work of Bao-an Li $et$ $al.$ \cite{li2025evolutions}, both $K_0$ and $F$ were constrained; while in the work of J. M. Wang $et$ $al.$ \cite{wang2024bayesian} and Yongjia Wang $et$ $al.$ \cite{Wang:2018hsw}, only $K_0$ was constrained. It can be seen that there are significant differences. The reasons for those differences might be: (1) These studies employ different observables, and different observables may reflect different density intervals of the EoS. (2) Different transport models are used in these literature. Different treatments of such as the mean field and collision terms may lead to different results. Although these results exhibit significant differences, several consistent patterns can be observed. For instance: The inverse correlation between $F$ and $K$ demonstrated in our study is also reported in Ref. \cite{li2025evolutions}. By considering the results from Refs. \cite{li2025evolutions,PhysRevC.110.064911}, the positive correlation between $F$ and $m^*$ found here in a much smaller range of values is supported.
%%%%%%%%%%%%%%%%%%%%%%%%%%%%%%%%%%%%%%%%%%%%%%%%%%%%%%%%%%%%%%%%%%
\section{Summary and outlook}\label{sec:4}

In summary, we have performed a statistically rigorous Bayesian analysis to extract posterior probability distributions 
for some of the parameters that characterize the properties of nuclear matter. FOPI experimental data, such as the directed flow $v_1$, elliptic flow $v_2$ and nuclear stopping power $vartl$ produced in $\rm ^{197}Au+^{197}Au$ at $E_{lab}$ = 0.25 and 0.4 GeV/nucleon, are used. Theoretical calculations are performed by the UrQMD model with considering different values of the incompressibility of the
nuclear equation of state $K_0$, the nucleon effective mass $m^*$, and the in-medium correction factor $F$. It is found that both $m^*$ and $F$ can be tightly constrained with the uncertainty $\le$ 15\%. However, obtaining a tight constraint on $K_0$ requires more effort. With the prior assumption of $K_0$ to be a Gaussian distribution centered at 240 MeV with a width of 60 MeV, we deduce $m^*/m_0 = 0.78^{+0.09}_{-0.10}$ and $F = 0.75^{+0.08}_{-0.07}$ with experimental data at $E_{lab}$ = 0.25 GeV/nucleon, and the obtained values have increased to $m^*/m_0 = 0.88^{+0.03}_{-0.03}$ and $F = 0.88^{+0.06}_{-0.07}$ at $E_{lab}$ = 0.4 GeV/nucleon. The obtained results are verified with the rapidity-dependent elliptic flow. 

In the present work, experimental data of the transverse-momentum-dependent flow are not used, as it is found that flow data at high transverse momentum region cannot be reproduced by calculations using the effective constant $F$, manifesting the importance of the density- and momentum-dependent in-medium correction factor in
nucleon-nucleon cross section when studying HICs at intermediate energies \cite{wei}. Including these transverse-momentum-dependent data is expected to lead to tighter constraints. Bayesian inference of more model parameters with more experimental data is being planned in our group.

%To achieve a tight constraint on the incompressibility of the nuclear equation of state, more sensitive observables and more precise experimental data are required. Furthermore, parameters in the initialization, potential and collision terms of transport models can also be set as free parameters in the Bayesian analysis, together with more precise experimental data coming from the upcoming CBM experiment at FAIR, and future experiments at HIAF and NICA, a more consistent and comprehensive understanding of the properties of nuclear matter can be expected in near future. 

%####################################################################

\section*{Acknowledgements}
We thank the anonymous referee for very useful comments and suggestions, which greatly improved this paper. We acknowledge fruitful discussions with the TMEP group. The authors are grateful to the C3S2 computing center in Huzhou University for calculation support. The work is supported by the National Natural Science Foundation of China (Nos. 12335008 and U2032145), the National Key R\&D Program of China (2023YFA1606402), the Foundation of National Key Laboratory of Plasma Physics (6142A04230203), Scientific Research Fund of Zhejiang Provincial Education Department (Y202353782), Huzhou Natural Science Foundation (Grant No. 2024YZ28), the Open Project of Guangxi Key Laboratory of Nuclear Physics and Nuclear Technology (No. NLK2023-02), and Central Government Guidance Funds
for Local Scientific and Technological Development, China
(No. Guike ZY22096024).

\bibliographystyle{elsarticle-num}
\bibliography{refer}  

\begin{thebibliography}{10}
\expandafter\ifx\csname url\endcsname\relax
  \def\url#1{\texttt{#1}}\fi
\expandafter\ifx\csname urlprefix\endcsname\relax\def\urlprefix{URL }\fi
\expandafter\ifx\csname href\endcsname\relax
  \def\href#1#2{#2} \def\path#1{#1}\fi

\bibitem{Drischler:2021kxf}
C.~Drischler, J.~W. Holt, C.~Wellenhofer, {Chiral Effective Field Theory and
  the High-Density Nuclear Equation of State}, Ann. Rev. Nucl. Part. Sci. 71
  (2021) 403--432.
\newblock \href {http://arxiv.org/abs/2101.01709} {\path{arXiv:2101.01709}},
  \href {https://doi.org/10.1146/annurev-nucl-102419-041903}
  {\path{doi:10.1146/annurev-nucl-102419-041903}}.

\bibitem{xu2019transport}
J.~Xu, {Transport approaches for the description of intermediate-energy
  heavy-ion collisions}, Prog. Part. Nucl. Phys. 106 (2019) 312--359.
\newblock \href {http://arxiv.org/abs/1904.00131} {\path{arXiv:1904.00131}},
  \href {https://doi.org/10.1016/j.ppnp.2019.02.009}
  {\path{doi:10.1016/j.ppnp.2019.02.009}}.

\bibitem{li2008recent}
B.~Li, L.~Chen, C.~M. Ko, {Recent Progress and New Challenges in Isospin
  Physics with Heavy-Ion Reactions}, Phys. Rept. 464 (2008) 113--281.
\newblock \href {http://arxiv.org/abs/0804.3580} {\path{arXiv:0804.3580}},
  \href {https://doi.org/10.1016/j.physrep.2008.04.005}
  {\path{doi:10.1016/j.physrep.2008.04.005}}.

\bibitem{Sorensen:2023zkk}
A.~Sorensen, et~al., {Dense nuclear matter equation of state from heavy-ion
  collisions}, Prog. Part. Nucl. Phys. 134 (2024) 104080.
\newblock \href {http://arxiv.org/abs/2301.13253} {\path{arXiv:2301.13253}},
  \href {https://doi.org/10.1016/j.ppnp.2023.104080}
  {\path{doi:10.1016/j.ppnp.2023.104080}}.

\bibitem{Huth:2021bsp}
S.~Huth, et~al., {Constraining Neutron-Star Matter with Microscopic and
  Macroscopic Collisions}, Nature 606 (2022) 276--280.
\newblock \href {http://arxiv.org/abs/2107.06229} {\path{arXiv:2107.06229}},
  \href {https://doi.org/10.1038/s41586-022-04750-w}
  {\path{doi:10.1038/s41586-022-04750-w}}.

\bibitem{SpiRIT:2023htl}
C.~Y. Tsang, et~al., {Constraining nucleon effective masses with flow and
  stopping observables from the S\ensuremath{\pi}RIT experiment}, Phys. Lett. B
  853 (2024) 138661.
\newblock \href {http://arxiv.org/abs/2312.06678} {\path{arXiv:2312.06678}},
  \href {https://doi.org/10.1016/j.physletb.2024.138661}
  {\path{doi:10.1016/j.physletb.2024.138661}}.

\bibitem{Tsang:2023vhh}
C.~Y. Tsang, M.~B. Tsang, W.~G. Lynch, R.~Kumar, C.~J. Horowitz, {Determination
  of the equation of state from nuclear experiments and neutron star
  observations}, Nature Astron. 8~(3) (2024) 328--336.
\newblock \href {http://arxiv.org/abs/2310.11588} {\path{arXiv:2310.11588}},
  \href {https://doi.org/10.1038/s41550-023-02161-z}
  {\path{doi:10.1038/s41550-023-02161-z}}.

\bibitem{doi:10.1126/science.1078070}
P.~Danielewicz, R.~Lacey, W.~G. Lynch,
  \href{https://www.science.org/doi/abs/10.1126/science.1078070}{Determination
  of the equation of state of dense matter}, Science 298~(5598) (2002)
  1592--1596.
\newblock \href
  {http://arxiv.org/abs/https://www.science.org/doi/pdf/10.1126/science.1078070}
  {\path{arXiv:https://www.science.org/doi/pdf/10.1126/science.1078070}}, \href
  {https://doi.org/10.1126/science.1078070}
  {\path{doi:10.1126/science.1078070}}.
\newline\urlprefix\url{https://www.science.org/doi/abs/10.1126/science.1078070}

\bibitem{TMEP:2022xjg}
H.~Wolter, et~al., {Transport model comparison studies of intermediate-energy
  heavy-ion collisions}, Prog. Part. Nucl. Phys. 125 (2022) 103962.
\newblock \href {http://arxiv.org/abs/2202.06672} {\path{arXiv:2202.06672}},
  \href {https://doi.org/10.1016/j.ppnp.2022.103962}
  {\path{doi:10.1016/j.ppnp.2022.103962}}.

\bibitem{garg2018compression}
U.~Garg, G.~Col\`o, {The compression-mode giant resonances and nuclear
  incompressibility}, Prog. Part. Nucl. Phys. 101 (2018) 55--95.
\newblock \href {http://arxiv.org/abs/1801.03672} {\path{arXiv:1801.03672}},
  \href {https://doi.org/10.1016/j.ppnp.2018.03.001}
  {\path{doi:10.1016/j.ppnp.2018.03.001}}.

\bibitem{xu2021bayesian}
J.~Xu, Z.~Zhang, B.-A. Li, {Bayesian uncertainty quantification for nuclear
  matter incompressibility}, Phys. Rev. C 104~(5) (2021) 054324.
\newblock \href {http://arxiv.org/abs/2107.10962} {\path{arXiv:2107.10962}},
  \href {https://doi.org/10.1103/PhysRevC.104.054324}
  {\path{doi:10.1103/PhysRevC.104.054324}}.

\bibitem{li2023toward}
Z.~Z. Li, Y.~F. Niu, G.~Col\`o, {Toward a Unified Description of Isoscalar
  Giant Monopole Resonances in a Self-Consistent Quasiparticle-Vibration
  Coupling Approach}, Phys. Rev. Lett. 131~(8) (2023) 082501.
\newblock \href {http://arxiv.org/abs/2211.01264} {\path{arXiv:2211.01264}},
  \href {https://doi.org/10.1103/PhysRevLett.131.082501}
  {\path{doi:10.1103/PhysRevLett.131.082501}}.

\bibitem{fuchs2001probing}
M.~Di~Toro, V.~Baran, M.~Colonna, V.~Greco, {Probing the Nuclear Symmetry
  Energy with Heavy Ion Collisions}, J. Phys. G 37 (2010) 083101.
\newblock \href {http://arxiv.org/abs/1003.2957} {\path{arXiv:1003.2957}},
  \href {https://doi.org/10.1088/0954-3899/37/8/083101}
  {\path{doi:10.1088/0954-3899/37/8/083101}}.

\bibitem{hartnack2006hadronic}
C.~Hartnack, H.~Oeschler, J.~Aichelin, Hadronic matter is soft, Physical review
  letters 96~(1) (2006) 012302.
\newblock \href {https://doi.org/10.1103/PhysRevLett.96.012302}
  {\path{doi:10.1103/PhysRevLett.96.012302}}.

\bibitem{sturm2001evidence}
C.~T. Sturm, et~al., {Evidence for a soft nuclear equation of state from kaon
  production in heavy ion collisions}, Phys. Rev. Lett. 86 (2001) 39--42.
\newblock \href {http://arxiv.org/abs/nucl-ex/0011001}
  {\path{arXiv:nucl-ex/0011001}}, \href
  {https://doi.org/10.1103/PhysRevLett.86.39}
  {\path{doi:10.1103/PhysRevLett.86.39}}.

\bibitem{Fuchs:2000kp}
C.~Fuchs, A.~Faessler, E.~Zabrodin, Y.-M. Zheng, {Probing the nuclear equation
  of state by K+ production in heavy ion collisions}, Phys. Rev. Lett. 86
  (2001) 1974--1977.
\newblock \href {http://arxiv.org/abs/nucl-th/0011102}
  {\path{arXiv:nucl-th/0011102}}, \href
  {https://doi.org/10.1103/PhysRevLett.86.1974}
  {\path{doi:10.1103/PhysRevLett.86.1974}}.

\bibitem{LeFevre:2015paj}
A.~Le~F\`evre, Y.~Leifels, W.~Reisdorf, J.~Aichelin, C.~Hartnack, {Constraining
  the nuclear matter equation of state around twice saturation density}, Nucl.
  Phys. A 945 (2016) 112--133.
\newblock \href {http://arxiv.org/abs/1501.05246} {\path{arXiv:1501.05246}},
  \href {https://doi.org/10.1016/j.nuclphysa.2015.09.015}
  {\path{doi:10.1016/j.nuclphysa.2015.09.015}}.

\bibitem{Wang:2018hsw}
Y.~Wang, C.~Guo, Q.~Li, A.~Le~F\`evre, Y.~Leifels, W.~Trautmann, {Determination
  of the nuclear incompressibility from the rapidity-dependent elliptic flow in
  heavy-ion collisions at beam energies 0.4 A \textendash{}1.0 A GeV}, Phys.
  Lett. B 778 (2018) 207--212.
\newblock \href {http://arxiv.org/abs/1804.04293} {\path{arXiv:1804.04293}},
  \href {https://doi.org/10.1016/j.physletb.2018.01.035}
  {\path{doi:10.1016/j.physletb.2018.01.035}}.

\bibitem{Oertel:2016bki}
M.~Oertel, M.~Hempel, T.~Kl\"ahn, S.~Typel, {Equations of state for supernovae
  and compact stars}, Rev. Mod. Phys. 89~(1) (2017) 015007.
\newblock \href {http://arxiv.org/abs/1610.03361} {\path{arXiv:1610.03361}},
  \href {https://doi.org/10.1103/RevModPhys.89.015007}
  {\path{doi:10.1103/RevModPhys.89.015007}}.

\bibitem{Pang:2022rzc}
P.~T.~H. Pang, et~al., {An updated nuclear-physics and multi-messenger
  astrophysics framework for binary neutron star mergers}, Nature Commun.
  14~(1) (2023) 8352.
\newblock \href {http://arxiv.org/abs/2205.08513} {\path{arXiv:2205.08513}},
  \href {https://doi.org/10.1038/s41467-023-43932-6}
  {\path{doi:10.1038/s41467-023-43932-6}}.

\bibitem{Koehn:2024set}
H.~Koehn, et~al., {From existing and new nuclear and astrophysical constraints
  to stringent limits on the equation of state of neutron-rich dense matter} (2
  2024).
\newblock \href {http://arxiv.org/abs/2402.04172} {\path{arXiv:2402.04172}}.

\bibitem{Zhang:2024npg}
N.-B. Zhang, B.-A. Li, {Impact of the nuclear equation of state on the
  formation of twin stars} (6 2024).
\newblock \href {http://arxiv.org/abs/2406.07396} {\path{arXiv:2406.07396}}.

\bibitem{Chatziioannou:2024tjq}
K.~Chatziioannou, H.~T. Cromartie, S.~Gandolfi, I.~Tews, D.~Radice, A.~W.
  Steiner, A.~L. Watts, {Neutron stars and the dense matter equation of state:
  from microscopic theory to macroscopic observations} (7 2024).
\newblock \href {http://arxiv.org/abs/2407.11153} {\path{arXiv:2407.11153}}.

\bibitem{Semposki:2024vnp}
A.~C. Semposki, C.~Drischler, R.~J. Furnstahl, J.~A. Melendez, D.~R. Phillips,
  {From chiral EFT to perturbative QCD: a Bayesian model mixing approach to
  symmetric nuclear matter} (4 2024).
\newblock \href {http://arxiv.org/abs/2404.06323} {\path{arXiv:2404.06323}}.

\bibitem{Sammarruca:2024rvc}
F.~Sammarruca, T.~Ajagbonna, {General features of the stellar matter equation
  of state from microscopic theory, new maximum-mass constraints, and
  causality} (12 2024).
\newblock \href {http://arxiv.org/abs/2501.00668} {\path{arXiv:2501.00668}}.

\bibitem{SpiRIT:2020sfn}
G.~Jhang, et~al., {Symmetry energy investigation with pion production from
  Sn+Sn systems}, Phys. Lett. B 813 (2021) 136016.
\newblock \href {http://arxiv.org/abs/2012.06976} {\path{arXiv:2012.06976}},
  \href {https://doi.org/10.1016/j.physletb.2020.136016}
  {\path{doi:10.1016/j.physletb.2020.136016}}.

\bibitem{TMEP:2016tup}
J.~Xu, et~al., {Understanding transport simulations of heavy-ion collisions at
  100A and 400A MeV: Comparison of heavy-ion transport codes under controlled
  conditions}, Phys. Rev. C 93~(4) (2016) 044609.
\newblock \href {http://arxiv.org/abs/1603.08149} {\path{arXiv:1603.08149}},
  \href {https://doi.org/10.1103/PhysRevC.93.044609}
  {\path{doi:10.1103/PhysRevC.93.044609}}.

\bibitem{TMEP:2017mex}
Y.-X. Zhang, et~al., {Comparison of heavy-ion transport simulations: Collision
  integral in a box}, Phys. Rev. C 97~(3) (2018) 034625.
\newblock \href {http://arxiv.org/abs/1711.05950} {\path{arXiv:1711.05950}},
  \href {https://doi.org/10.1103/PhysRevC.97.034625}
  {\path{doi:10.1103/PhysRevC.97.034625}}.

\bibitem{TMEP:2019yci}
A.~Ono, et~al., {Comparison of heavy-ion transport simulations: Collision
  integral with pions and \ensuremath{\Delta} resonances in a box}, Phys. Rev.
  C 100~(4) (2019) 044617.
\newblock \href {http://arxiv.org/abs/1904.02888} {\path{arXiv:1904.02888}},
  \href {https://doi.org/10.1103/PhysRevC.100.044617}
  {\path{doi:10.1103/PhysRevC.100.044617}}.

\bibitem{TMEP:2021ljz}
M.~Colonna, et~al., {Comparison of heavy-ion transport simulations: Mean-field
  dynamics in a box}, Phys. Rev. C 104~(2) (2021) 024603.
\newblock \href {http://arxiv.org/abs/2106.12287} {\path{arXiv:2106.12287}},
  \href {https://doi.org/10.1103/PhysRevC.104.024603}
  {\path{doi:10.1103/PhysRevC.104.024603}}.

\bibitem{Gao:2022shr}
B.~Gao, Y.~Wang, Z.~Gao, Q.~Li, {Elliptic flow in heavy-ion collisions at
  intermediate energy: The role of impact parameter, mean field potential, and
  collision term}, Phys. Lett. B 838 (2023) 137685.
\newblock \href {http://arxiv.org/abs/2210.08213} {\path{arXiv:2210.08213}},
  \href {https://doi.org/10.1016/j.physletb.2023.137685}
  {\path{doi:10.1016/j.physletb.2023.137685}}.

\bibitem{Wang:2024ktk}
Y.~Wang, B.~Gao, G.~Wei, P.~Li, Q.~Li, {Time evolution of elliptic flow and
  medium density in heavy-ion collisions at intermediate energies}, Phys. Rev.
  C 110~(4) (2024) 044606.
\newblock \href {https://doi.org/10.1103/PhysRevC.110.044606}
  {\path{doi:10.1103/PhysRevC.110.044606}}.

\bibitem{liu2023impacts}
Y.-Y. Liu, J.-P. Yang, Y.-J. Wang, Q.-F. Li, Z.-X. Li, C.-J. Xia, Y.-X. Zhang,
  \href{https://doi.org/10.1007/s41365-024-01607-x}{A perspective on describing
  nucleonic flow and pionic observables within the ultra-relativistic quantum
  molecular dynamics model}, Nuclear Science and Techniques 36~(3) (2025) 45.
\newblock \href {https://doi.org/10.1007/s41365-024-01607-x}
  {\path{doi:10.1007/s41365-024-01607-x}}.
\newline\urlprefix\url{https://doi.org/10.1007/s41365-024-01607-x}

\bibitem{Andronic:2006ra}
A.~Andronic, J.~Lukasik, W.~Reisdorf, W.~Trautmann, {Systematics of Stopping
  and Flow in Au+Au Collisions}, Eur. Phys. J. A 30 (2006) 31--46.
\newblock \href {http://arxiv.org/abs/nucl-ex/0608015}
  {\path{arXiv:nucl-ex/0608015}}, \href
  {https://doi.org/10.1140/epja/i2006-10101-2}
  {\path{doi:10.1140/epja/i2006-10101-2}}.

\bibitem{HADES:2020ver}
J.~Adamczewski-Musch, et~al., {Charged-pion production in $\mathbf {Au+Au}$
  collisions at $\sqrt{\mathbf {s}_{\mathbf {NN}}} = 2.4~{\mathbf {GeV}}$:
  HADES Collaboration}, Eur. Phys. J. A 56~(10) (2020) 259.
\newblock \href {http://arxiv.org/abs/2005.08774} {\path{arXiv:2005.08774}},
  \href {https://doi.org/10.1140/epja/s10050-020-00237-2}
  {\path{doi:10.1140/epja/s10050-020-00237-2}}.

\bibitem{Mamamev:2024ynl}
M.~Mamamev, A.~Taranenko, A.~Demanov, P.~Parfenov, V.~Troshin, {Analysis Note:
  Directed flow $v_1$ of protons in the Xe+Cs(I) collisions at 3.8 AGeV} (12
  2024).
\newblock \href {http://arxiv.org/abs/2412.08570} {\path{arXiv:2412.08570}}.

\bibitem{Sharma:2023tvv}
S.~R. Sharma, {First-Order Event Plane Correlated Directed and Triangular Flow
  from Fixed-Target Energies at RHIC-STAR}, Universe 10~(3) (2024) 118.
\newblock \href {http://arxiv.org/abs/2312.02666} {\path{arXiv:2312.02666}},
  \href {https://doi.org/10.3390/universe10030118}
  {\path{doi:10.3390/universe10030118}}.

\bibitem{ono2019dynamics}
A.~Ono, {Dynamics of clusters and fragments in heavy-ion collisions}, Prog.
  Part. Nucl. Phys. 105 (2019) 139--179.
\newblock \href {http://arxiv.org/abs/1903.00608} {\path{arXiv:1903.00608}},
  \href {https://doi.org/10.1016/j.ppnp.2018.11.001}
  {\path{doi:10.1016/j.ppnp.2018.11.001}}.

\bibitem{bleicher2022modelling}
M.~Bleicher, E.~Bratkovskaya, {Modelling relativistic heavy-ion collisions with
  dynamical transport approaches}, Prog. Part. Nucl. Phys. 122 (2022) 103920.
\newblock \href {https://doi.org/10.1016/j.ppnp.2021.103920}
  {\path{doi:10.1016/j.ppnp.2021.103920}}.

\bibitem{li2018nucleon}
B.~Li, B.~Cai, L.~Chen, J.~Xu, {Nucleon Effective Masses in Neutron-Rich
  Matter}, Prog. Part. Nucl. Phys. 99 (2018) 29--119.
\newblock \href {http://arxiv.org/abs/1801.01213} {\path{arXiv:1801.01213}},
  \href {https://doi.org/10.1016/j.ppnp.2018.01.001}
  {\path{doi:10.1016/j.ppnp.2018.01.001}}.

\bibitem{Han:2022quc}
S.~C. Han, X.~L. Shang, W.~Zuo, G.~C. Yong, Y.~Gao, {In-medium nucleon-nucleon
  cross section~in nuclear matter}, Phys. Rev. C 106~(6) (2022) 064332.
\newblock \href {https://doi.org/10.1103/PhysRevC.106.064332}
  {\path{doi:10.1103/PhysRevC.106.064332}}.

\bibitem{Cui:2018dex}
Y.~Cui, Y.~Zhang, Z.~Li, {In-medium $\textrm{NN}\to \textrm{N}\Delta$ cross
  section and its dependence on effective Lagrange parameters in
  isospin-asymmetric nuclear matter}, Chin. Phys. C 43~(2) (2019) 024105.
\newblock \href {http://arxiv.org/abs/1801.05960} {\path{arXiv:1801.05960}},
  \href {https://doi.org/10.1088/1674-1137/43/2/024105}
  {\path{doi:10.1088/1674-1137/43/2/024105}}.

\bibitem{song2023medium}
Y.~D. Song, R.~Wang, Z.~Zhang, Y.~G. Ma, {In-medium nucleon-nucleon cross
  sections~from characteristics of nuclear giant resonances and nuclear
  stopping power}, Phys. Rev. C 108~(6) (2023) 064603.
\newblock \href {https://doi.org/10.1103/PhysRevC.108.064603}
  {\path{doi:10.1103/PhysRevC.108.064603}}.

\bibitem{li2023bayesian}
B.~Li, W.~Xie, {Bayesian inference of in-medium baryon-baryon scattering cross
  sections from HADES proton flow data}, Nucl. Phys. A 1039 (2023) 122726.
\newblock \href {http://arxiv.org/abs/2303.10474} {\path{arXiv:2303.10474}},
  \href {https://doi.org/10.1016/j.nuclphysa.2023.122726}
  {\path{doi:10.1016/j.nuclphysa.2023.122726}}.

\bibitem{li2025evolutions}
B.-A. Li, W.-J. Xie, {Evolution of in-medium baryon-baryon scattering cross
  sections~and stiffness of dense nuclear matter from Bayesian analyses of FOPI
  proton-flow excitation functions}, Phys. Rev. C 111~(5) (2025) 054602.
\newblock \href {http://arxiv.org/abs/2501.02579} {\path{arXiv:2501.02579}},
  \href {https://doi.org/10.1103/PhysRevC.111.054602}
  {\path{doi:10.1103/PhysRevC.111.054602}}.

\bibitem{gal2022bayesian}
Y.~Gal, P.~Koumoutsakos, F.~Lanusse, G.~Louppe, C.~Papadimitriou, Bayesian
  uncertainty quantification for machine-learned models in physics, Nature
  Reviews Physics 4~(9) (2022) 573--577.
\newblock \href {https://doi.org/10.1038/s42254-022-00498-4}
  {\path{doi:10.1038/s42254-022-00498-4}}.

\bibitem{PhysRevC.110.064911}
M.~D. Cozma, Equation of state of nuclear matter from collective flows and
  stopping in intermediate-energy heavy-ion collisions, Phys. Rev. C 110 (2024)
  064911.
\newblock \href {https://doi.org/10.1103/PhysRevC.110.064911}
  {\path{doi:10.1103/PhysRevC.110.064911}}.

\bibitem{kuttan2022qcd}
M.~Omana~Kuttan, J.~Steinheimer, K.~Zhou, H.~Stoecker, {QCD Equation of State
  of Dense Nuclear Matter from a Bayesian Analysis of Heavy-Ion Collision
  Data}, Phys. Rev. Lett. 131~(20) (2023) 202303.
\newblock \href {http://arxiv.org/abs/2211.11670} {\path{arXiv:2211.11670}},
  \href {https://doi.org/10.1103/PhysRevLett.131.202303}
  {\path{doi:10.1103/PhysRevLett.131.202303}}.

\bibitem{Bernhard:2018hnz}
J.~E. Bernhard, {Bayesian parameter estimation for relativistic heavy-ion
  collisions}, Ph.D. thesis, Duke U. (4 2018).
\newblock \href {http://arxiv.org/abs/1804.06469} {\path{arXiv:1804.06469}}.

\bibitem{wang2024bayesian}
J.~M. Wang, X.~G. Deng, W.~J. Xie, B.~A. Li, Y.~G. Ma, {Bayesian inference of
  nuclear incompressibility from proton elliptic flow in central Au+Au
  collisions at 400 MeV/nucleon} (6 2024).
\newblock \href {http://arxiv.org/abs/2406.07051} {\path{arXiv:2406.07051}}.

\bibitem{reisdorf1997collective}
W.~Reisdorf, H.~Ritter, Collective flow in heavy-ion collisions, Annual Review
  of Nuclear and Particle Science 47~(1) (1997) 663--709.
\newblock \href {https://doi.org/10.1146/annurev.nucl.47.1.663}
  {\path{doi:10.1146/annurev.nucl.47.1.663}}.

\bibitem{herrmann1999collective}
N.~Herrmann, J.~P. Wessels, T.~Wienold, Collective flow in heavy-ion
  collisions, Annual Review of Nuclear and Particle Science 49~(1) (1999)
  581--632.
\newblock \href {https://doi.org/10.1146/annurev.nucl.47.1.663}
  {\path{doi:10.1146/annurev.nucl.47.1.663}}.

\bibitem{heinz2013collective}
U.~Heinz, R.~Snellings, {Collective flow and viscosity in relativistic
  heavy-ion collisions}, Ann. Rev. Nucl. Part. Sci. 63 (2013) 123--151.
\newblock \href {http://arxiv.org/abs/1301.2826} {\path{arXiv:1301.2826}},
  \href {https://doi.org/10.1146/annurev-nucl-102212-170540}
  {\path{doi:10.1146/annurev-nucl-102212-170540}}.

\bibitem{Lan2022}
S.-W. Lan, S.-S. Shi, Anisotropic flow in high baryon density region, Nuclear
  Science and Techniques 33~(2) (2022) 21.

\bibitem{elfner2023exploration}
H.~Elfner, B.~M\"uller, {The exploration of hot and dense nuclear matter:
  introduction to relativistic heavy-ion physics}, J. Phys. G 50~(10) (2023)
  103001.
\newblock \href {http://arxiv.org/abs/2210.12056} {\path{arXiv:2210.12056}},
  \href {https://doi.org/10.1088/1361-6471/ace824}
  {\path{doi:10.1088/1361-6471/ace824}}.

\bibitem{Wang:2020dru}
Y.~Wang, Q.~Li, Y.~Leifels, A.~Le~F\`evre, {Study of the nuclear symmetry
  energy from the rapidity-dependent elliptic flow in heavy-ion collisions
  around 1 GeV/nucleon regime}, Phys. Lett. B 802 (2020) 135249.
\newblock \href {https://doi.org/10.1016/j.physletb.2020.135249}
  {\path{doi:10.1016/j.physletb.2020.135249}}.

\bibitem{FOPI:2011aa}
W.~Reisdorf, et~al., {Systematics of azimuthal asymmetries in heavy ion
  collisions in the 1 A GeV regime}, Nucl. Phys. A 876 (2012) 1--60.
\newblock \href {http://arxiv.org/abs/1112.3180} {\path{arXiv:1112.3180}},
  \href {https://doi.org/10.1016/j.nuclphysa.2011.12.006}
  {\path{doi:10.1016/j.nuclphysa.2011.12.006}}.

\bibitem{PhysRevLett.92.232301}
W.~Reisdorf, A.~Andronic, A.~Gobbi, O.~N. Hartmann, N.~Herrmann, K.~D.
  Hildenbrand, Y.~J. Kim, M.~Kirejczyk, P.~Koczo\ifmmode~\acute{n}\else
  \'{n}\fi{}, T.~Kress, Y.~Leifels, A.~Sch\"uttauf,
  Z.~Tymi\ifmmode~\acute{n}\else \'{n}\fi{}ski, Z.~G. Xiao, J.~P. Alard,
  V.~Barret, Z.~Basrak, N.~Bastid, M.~L. Benabderrahmane,
  R.~\ifmmode~\check{C}\else \v{C}\fi{}aplar, P.~Crochet, P.~Dupieux,
  M.~D\ifmmode~\check{z}\else \v{z}\fi{}elalija, Z.~Fodor, Y.~Grishkin,
  B.~Hong, J.~Kecskemeti, M.~Korolija, R.~Kotte, A.~Lebedev, X.~Lopez,
  M.~Merschmeyer, J.~M\"osner, W.~Neubert, D.~Pelte, M.~Petrovici, F.~Rami,
  B.~de~Schauenburg, Z.~Seres, B.~Sikora, K.~S. Sim, V.~Simion,
  K.~Siwek-Wilczy\ifmmode~\acute{n}\else \'{n}\fi{}ska, V.~Smolyankin,
  M.~Stockmeier, G.~Stoicea, P.~Wagner, K.~Wi\ifmmode~\acute{s}\else
  \'{s}\fi{}niewski, D.~Wohlfarth, I.~Yushmanov, A.~Zhilin,
  \href{https://link.aps.org/doi/10.1103/PhysRevLett.92.232301}{Nuclear
  stopping from $0.09a$ to $1.93a\text{ }\text{ }\mathrm{GeV}$ and its
  correlation to flow}, Phys. Rev. Lett. 92 (2004) 232301.
\newblock \href {https://doi.org/10.1103/PhysRevLett.92.232301}
  {\path{doi:10.1103/PhysRevLett.92.232301}}.
\newline\urlprefix\url{https://link.aps.org/doi/10.1103/PhysRevLett.92.232301}

\bibitem{lehaut2010study}
G.~Lehaut, et~al., {Study of Nuclear Stopping in Central Collisions at
  Intermediate Energies}, Phys. Rev. Lett. 104 (2010) 232701.
\newblock \href {https://doi.org/10.1103/PhysRevLett.104.232701}
  {\path{doi:10.1103/PhysRevLett.104.232701}}.

\bibitem{Lopez:2014dga}
O.~Lopez, et~al., {In-medium effects for nuclear matter in the Fermi energy
  domain}, Phys. Rev. C 90~(6) (2014) 064602, [Erratum: Phys.Rev.C 90, 069903
  (2014)].
\newblock \href {http://arxiv.org/abs/1409.0735} {\path{arXiv:1409.0735}},
  \href {https://doi.org/10.1103/PhysRevC.90.064602}
  {\path{doi:10.1103/PhysRevC.90.064602}}.

\bibitem{FOPI:2010xrt}
W.~Reisdorf, et~al., {Systematics of central heavy ion collisions in the 1A GeV
  regime}, Nucl. Phys. A 848 (2010) 366--427.
\newblock \href {http://arxiv.org/abs/1005.3418} {\path{arXiv:1005.3418}},
  \href {https://doi.org/10.1016/j.nuclphysa.2010.09.008}
  {\path{doi:10.1016/j.nuclphysa.2010.09.008}}.

\bibitem{Li11}
Q.~Li, C.~Shen, C.~Guo, Y.~Wang, Z.~Li, J.~Lukasik, W.~Trautmann,
  {Nonequilibrium dynamics in heavy-ion collisions at low energies available at
  the GSI Schwerionen Synchrotron}, Phys. Rev. C 83 (2011) 044617.
\newblock \href {https://doi.org/10.1103/PhysRevC.83.044617}
  {\path{doi:10.1103/PhysRevC.83.044617}}.

\bibitem{wang2014collective}
Y.~Wang, C.~Guo, Q.~Li, H.~Zhang, Z.~Li, W.~Trautmann, {Collective flows of
  light particles in the Au+Au collisions at intermediate energies}, Phys. Rev.
  C 89~(3) (2014) 034606.
\newblock \href {http://arxiv.org/abs/1305.4730} {\path{arXiv:1305.4730}},
  \href {https://doi.org/10.1103/PhysRevC.89.034606}
  {\path{doi:10.1103/PhysRevC.89.034606}}.

\bibitem{wang2020application}
Y.~Wang, Q.~Li, {Application of microscopic transport model in the study of
  nuclear equation of state from heavy ion collisions at intermediate
  energies}, Front. Phys. (Beijing) 15~(4) (2020) 44302.
\newblock \href {https://doi.org/10.1007/s11467-020-0964-6}
  {\path{doi:10.1007/s11467-020-0964-6}}.

\bibitem{Li:2018wpv}
P.~Li, Y.~Wang, Q.~Li, C.~Guo, H.~Zhang, {Effects of the in-medium
  nucleon-nucleon cross section on collective flow and nuclear stopping in
  heavy-ion collisions in the Fermi-energy domain}, Phys. Rev. C 97~(4) (2018)
  044620.
\newblock \href {http://arxiv.org/abs/1804.04288} {\path{arXiv:1804.04288}},
  \href {https://doi.org/10.1103/PhysRevC.97.044620}
  {\path{doi:10.1103/PhysRevC.97.044620}}.

\bibitem{danielewicz2000determination}
P.~Danielewicz, {Determination of the mean field momentum dependence using
  elliptic flow}, Nucl. Phys. A 673 (2000) 375--410.
\newblock \href {http://arxiv.org/abs/nucl-th/9912027}
  {\path{arXiv:nucl-th/9912027}}, \href
  {https://doi.org/10.1016/S0375-9474(00)00083-X}
  {\path{doi:10.1016/S0375-9474(00)00083-X}}.

\bibitem{li2022accessing}
P.~Li, Y.~Wang, Q.~Li, H.~Zhang, {Accessing the in-medium effects on
  nucleon-nucleon elastic cross section with collective flows and nuclear
  stopping}, Phys. Lett. B 828 (2022) 137019.
\newblock \href {http://arxiv.org/abs/2203.05855} {\path{arXiv:2203.05855}},
  \href {https://doi.org/10.1016/j.physletb.2022.137019}
  {\path{doi:10.1016/j.physletb.2022.137019}}.

\bibitem{hartnack2012strangeness}
C.~Hartnack, H.~Oeschler, Y.~Leifels, E.~L. Bratkovskaya, J.~Aichelin,
  {Strangeness Production close to Threshold in Proton-Nucleus and Heavy-Ion
  Collisions}, Phys. Rept. 510 (2012) 119--200.
\newblock \href {http://arxiv.org/abs/1106.2083} {\path{arXiv:1106.2083}},
  \href {https://doi.org/10.1016/j.physrep.2011.08.004}
  {\path{doi:10.1016/j.physrep.2011.08.004}}.

\bibitem{le2018origin}
A.~Le~F\`evre, Y.~Leifels, C.~Hartnack, J.~Aichelin, {Origin of elliptic flow
  and its dependence on the equation of state in heavy ion reactions at
  intermediate energies}, Phys. Rev. C 98~(3) (2018) 034901.
\newblock \href {http://arxiv.org/abs/1611.07500} {\path{arXiv:1611.07500}},
  \href {https://doi.org/10.1103/PhysRevC.98.034901}
  {\path{doi:10.1103/PhysRevC.98.034901}}.

\bibitem{Du:2023ype}
H.~Du, G.-F. Wei, G.-C. Yong, {Directed and elliptic flows of protons and
  deuterons in HADES Au+Au collisions at sNN=2.4 GeV}, Phys. Lett. B 839 (2023)
  137823.
\newblock \href {http://arxiv.org/abs/2302.07037} {\path{arXiv:2302.07037}},
  \href {https://doi.org/10.1016/j.physletb.2023.137823}
  {\path{doi:10.1016/j.physletb.2023.137823}}.

\bibitem{Kireyeu:2024hjo}
V.~Kireyeu, V.~Voronyuk, M.~Winn, S.~Gl\"a\ss{}el, J.~Aichelin, C.~Blume,
  E.~Bratkovskaya, G.~Coci, J.~Zhao, {Constraints on the equation-of-state from
  low energy heavy-ion collisions within the PHQMD microscopic approach with
  momentum-dependent potential} (11 2024).
\newblock \href {http://arxiv.org/abs/2411.04969} {\path{arXiv:2411.04969}}.

\bibitem{Li:2018bus}
P.~Li, Y.~Wang, Q.~Li, H.~Zhang, {Collective flow and nuclear stopping in heavy
  ion collisions in Fermi energy domain}, Nucl. Sci. Tech. 29~(12) (2018) 177.
\newblock \href {https://doi.org/10.1007/s41365-018-0510-1}
  {\path{doi:10.1007/s41365-018-0510-1}}.

\bibitem{barker2019shear}
X.~Deng, D.~Fang, Y.~Ma, {Shear viscosity of nucleonic matter}, Prog. Part.
  Nucl. Phys. 136 (2024) 104095.
\newblock \href {http://arxiv.org/abs/2401.02293} {\path{arXiv:2401.02293}},
  \href {https://doi.org/10.1016/j.ppnp.2023.104095}
  {\path{doi:10.1016/j.ppnp.2023.104095}}.

\bibitem{isse2005mean}
M.~Isse, A.~Ohnishi, N.~Otuka, P.~K. Sahu, Y.~Nara, {Mean-field effects on
  collective flows in high-energy heavy-ion collisions from AGS to SPS
  energies}, Phys. Rev. C 72 (2005) 064908.
\newblock \href {http://arxiv.org/abs/nucl-th/0502058}
  {\path{arXiv:nucl-th/0502058}}, \href
  {https://doi.org/10.1103/PhysRevC.72.064908}
  {\path{doi:10.1103/PhysRevC.72.064908}}.

\bibitem{andronic2003directed}
A.~Andronic, et~al., {Directed flow in Au + Au, Xe + CsI and Ni + Ni collisions
  and the nuclear equation of state}, Phys. Rev. C 67 (2003) 034907.
\newblock \href {http://arxiv.org/abs/nucl-ex/0301009}
  {\path{arXiv:nucl-ex/0301009}}, \href
  {https://doi.org/10.1103/PhysRevC.67.034907}
  {\path{doi:10.1103/PhysRevC.67.034907}}.

\bibitem{wei}
G.~Wei, et~al., {to be submitted}.

\bibitem{yang2020bayesian}
L.~Yang, et~al., {Bayesian analysis on interactions of exotic nuclear systems},
  Phys. Lett. B 807 (2020) 135540.
\newblock \href {https://doi.org/10.1016/j.physletb.2020.135540}
  {\path{doi:10.1016/j.physletb.2020.135540}}.

\bibitem{mantysaari2022bayesian}
H.~M\"antysaari, B.~Schenke, C.~Shen, W.~Zhao, {Bayesian inference of the
  fluctuating proton shape}, Phys. Lett. B 833 (2022) 137348.
\newblock \href {http://arxiv.org/abs/2202.01998} {\path{arXiv:2202.01998}},
  \href {https://doi.org/10.1016/j.physletb.2022.137348}
  {\path{doi:10.1016/j.physletb.2022.137348}}.

\bibitem{everett2021multisystem}
D.~Everett, et~al., {Multisystem Bayesian constraints on the transport
  coefficients of QCD matter}, Phys. Rev. C 103~(5) (2021) 054904.
\newblock \href {http://arxiv.org/abs/2011.01430} {\path{arXiv:2011.01430}},
  \href {https://doi.org/10.1103/PhysRevC.103.054904}
  {\path{doi:10.1103/PhysRevC.103.054904}}.

\bibitem{Wang:2023kcg}
Y.~Wang, Q.~Li, {Machine learning transforms the inference of the nuclear
  equation of state}, Front. Phys. (Beijing) 18~(6) (2023) 64402.
\newblock \href {http://arxiv.org/abs/2305.16686} {\path{arXiv:2305.16686}},
  \href {https://doi.org/10.1007/s11467-023-1313-3}
  {\path{doi:10.1007/s11467-023-1313-3}}.

\bibitem{semposki2022interpolating}
A.~C. Semposki, R.~J. Furnstahl, D.~R. Phillips, {Interpolating between small-
  and large-g expansions using Bayesian model mixing}, Phys. Rev. C 106~(4)
  (2022) 044002.
\newblock \href {http://arxiv.org/abs/2206.04116} {\path{arXiv:2206.04116}},
  \href {https://doi.org/10.1103/PhysRevC.106.044002}
  {\path{doi:10.1103/PhysRevC.106.044002}}.

\end{thebibliography}

% 附录开始
\appendix
\section*{Appendix} % 添加不带编号的附录总标题
\addcontentsline{toc}{section}{Appendix} % 手动添加到目录
\renewcommand{\thesection}{\Alph{section}} % 设置附录编号为A, B, C...

\begin{figure}[!htbp]
   \centering
    \includegraphics[width=0.5\textwidth]{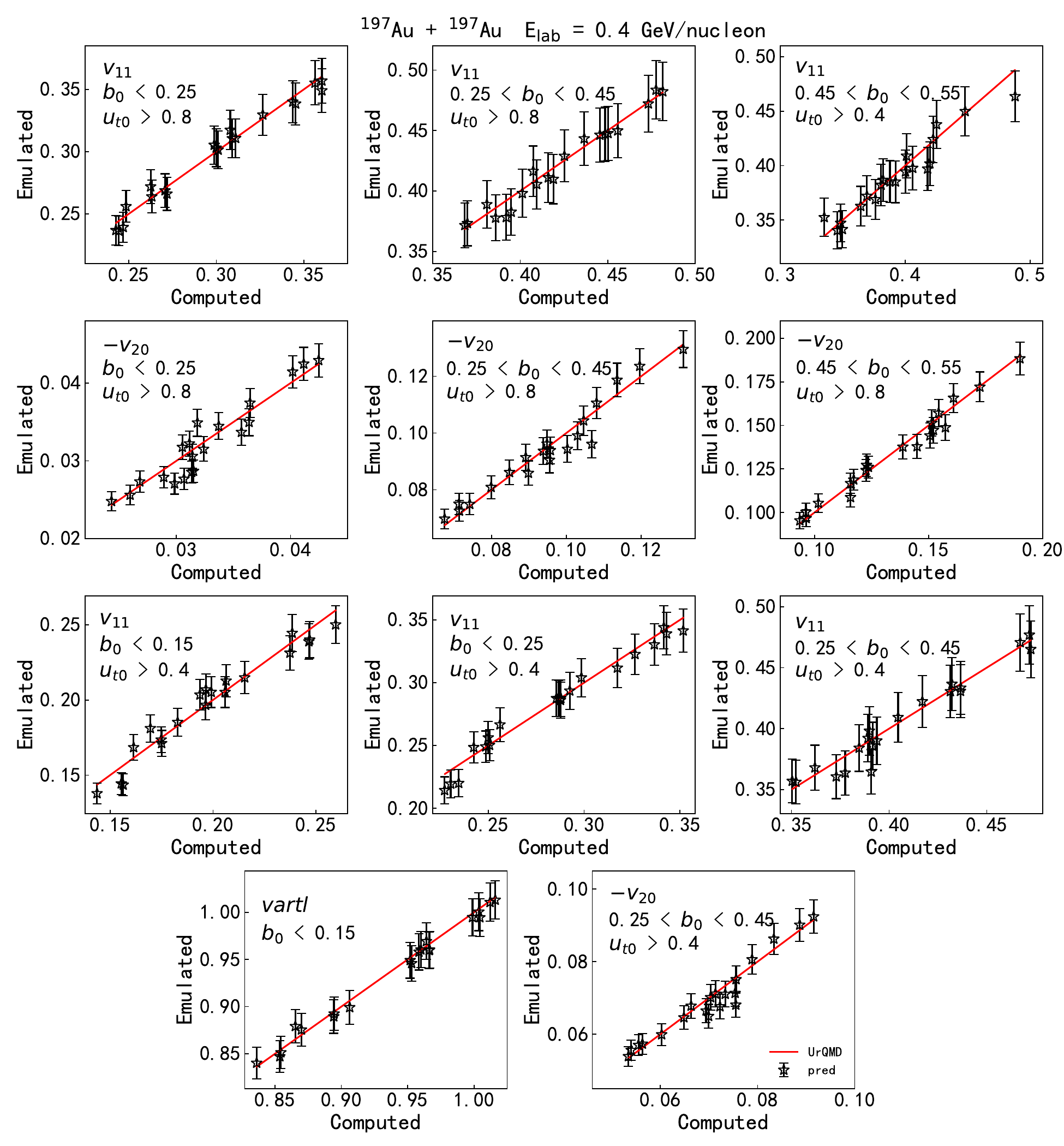}
    \caption{Emulated vs computed for all observables considered. Successful emulation is clustered around $y = x$.}
   \label{fig:2}
\end{figure}

The canonical choice for model emulators are Gaussian processes, many studies have shown that Gaussian processes are sufficiently flexible to emulate a wide variety of models \cite{SpiRIT:2023htl,yang2020bayesian,mantysaari2022bayesian,li2025evolutions,li2023bayesian}. Thus, a Gaussian process simulator is used in this work, and the results from the calculations with the 150 parameter sets are fed to the Gaussian process to train the emulators. In addition, 20 parameter sets of the UrQMD model with randomly chosen $K_0$, $m^*$, and $F$ are run and the results on observables are used to test the performance of the emulators. Gaussian process models are trained for each observables. The performance of the emulators in predicting observables are shown in Fig. \ref{fig:2}. As can be seen, the results obtained from emulators are in line with the one obtained from UrQMD model. Hence, results from emulators can be used during the Bayesian analysis. 

Bayes' theorem is a general and systematic method to constrain the probability distribution of model parameters $\boldsymbol{\theta}$ by comparing model calculations ($y_{\cal}$) with experimental measurements $y_{\exp}$. It is written in the following form \cite{everett2021multisystem,Wang:2023kcg}:

\begin{equation}
P\left(\boldsymbol{\theta} \mid y_{\exp }\right) \propto P\left(y_{\exp } \mid \boldsymbol{\theta}\right) P(\boldsymbol{\theta}),
\end{equation}

\begin{equation}
P\left(y_{\exp } \mid \boldsymbol{\theta}\right) = \exp [-\frac{1}{2}\left(\boldsymbol{\theta}-y_{\exp }\right)^{\top} \Sigma^{-1}\left(\boldsymbol{\theta}-y_{\exp }\right)],
\end{equation}

\begin{equation}
\Sigma=\operatorname{diag}\left(\sqrt{\sigma_{exp, 1}^{2}+\sigma_{emu, 1}^{2}}  , \cdots, \sqrt{\sigma_{exp, n}^{2}+\sigma_{emu, n}^{2}}\right).
\end{equation}

Here, $P(\boldsymbol{\theta})$ represents the prior distribution, which encodes the prior information about the parameters. $P\left(y_{\exp} \mid \boldsymbol{\theta}\right)$ denotes the likelihood function, specifying how well a given set of parameters describes the observables. In this work, a Gaussian form function that combines both the experimental error and the Gaussian process error is used. $\sigma_{\text {exp}, i}$ and $\sigma_{\text {emu}, i}$ are the experimental error and emulation error of the i-th observable, respectively. $P\left(\boldsymbol{\theta}\mid y_{\exp }\right)$ is the posterior distribution, which shows the most likely parameter distribution based on the given experimental data. To obtain $P\left(\boldsymbol{\theta}\mid y_{\exp }\right)$, the Markov chain Monte
Carlo (MCMC) method is used to efficiently sample the phase space of the parameters \cite{semposki2022interpolating,mantysaari2022bayesian}.

\end{document}